\documentstyle[aps,twocolumn,epsf]{revtex}
\begin{document}
\draft
 \title{
Metal-insulator transition in disordered 2DEG\\
 including temperature effects.
}
\author{J.S. Thakur$^\star$, Lerwen Liu$^\dagger$ and D.
Neilson$^\star$$^\dagger$} 
\address{
$^\star$School of Physics, The University of New South Wales, 
Sydney 2052, Australia\\
$^\dagger$Scuola Normale Superiore, Piazza dei Cavalieri 7, 56126
Pisa, Italy
\\[3pt]
(Submitted to Phys.\ Rev.\ B )
\  \\ \medskip}\author{\small\parbox{14cm}{\small
We calculate self-consistently the mutual dependence of electron
correlations and electron-defect scattering for a two dimensional
electron gas at finite temperature. We employ an STLS approach to
calculate the electron correlations while the electron scattering rate
off Coulombic impurities and surface roughness is calculated using
self-consistent current-relaxation theory.  The methods are combined
and self-consistently solved.  We discuss a metal-insulator transition
for a range of disorder levels and electron densities.  Our results are
in good agreement with recent experimental observations.
 \\[3pt]{PACS numbers:
73.20.Dx,71.30.+h,71.55.-i}
}}\address{} \maketitle

\narrowtext

In recent experiments \cite{Kravchenko,pudalov,simmons} on
two-dimensional electronic systems in zero magnetic field a well
defined metal-insulator transition has been observed. The transition
contradicts the prediction \cite{AALR} that two dimensional electron
systems are always localized in the presence of any disorder.  The 
mobility is large at the transition point ($\mu>1$m$^2$/Vs).  For such
a high mobility the scattering off defects will be weak and electron
correlations could make a significant difference.

Both disorder and correlations by themselves can lead to at least two
different types of localization.  In the presence of disorder Abrahams
{\it et al} \cite{AALR} showed that non-interacting electrons in two
dimensions cannot sustain static conductivity no matter how small the
level of disorder is. On the other hand if there are strong
correlations between the charge carriers this can cause a different
type of localization in disorder-free systems which is associated with
Wigner crystallization \cite{Wigner}.  In real systems there are both
electron-electron and electron-impurity interactions.  With weak
disorder and in the low electron density limit the electron-electron
interactions should dominate leading to Wigner crystallization, while
in the high electron density limit the electron correlations are very
weak and the localization should be of the Anderson type.   Between
these two extremes disorder and correlations compete with each other
to decide the nature of the localization.

In this paper we examine the interdependence of correlations and defect
scattering at finite temperatures.   We use the self-consistent
formalism of Singwi, Tosi, Land and Sj\"olander (STLS) \cite{STLS} to
treat electron-electron correlations, while for the electron-defect
scattering we use a memory function approach \cite{Gotze} to calculate
the decay time of the density fluctuations from scatterings off
defects.  Previous calculations of the influence of disorder have
either used the first Born approximation or else they have introduced
an adjustable parameter $\gamma$ for the electron scattering rate from
the disorder.  In these approaches the scattering rate is not affected
by correlations while the two effects are in fact interdependent and
should be self-consistently linked.

We consider a random distribution of Coulombic impurities with charge
$e$ and density $n_i$.  The charge carrier (electron or hole) density
$n_c$ is chosen low enough that the carriers occupy only the lowest
energy sub-band.  In the STLS formalism the bare Coulomb potential
$V(q)=2\pi e^2/{q\epsilon}$ acting between the carriers is replaced by
an effective interaction $V_{\text{eff}}(q)=V(q)[1-G(q)]$, where $G(q)$
is a local field factor.  $V_{\text{eff}}(q)$ is generally weaker than
$V(q)$ because of the correlations between the carriers.  The response
function becomes
 \begin{equation}
 \chi(q,\omega)=\frac{\chi^{(0)}(q, 
 \omega)}{1+V(q)\left[1-G(q)\right]\chi^{(0)}(q,\omega)}\ .
 \label{chiSTLS} 
 \end{equation} 

We include the effect of disorder by replacing the free particle
dynamical susceptibility $\chi^{(0)}(q,\omega)$ in Eq.\  \ref{chiSTLS}
by the response function for non-interacting charge carriers scattering
from the disorder \cite{Mermin},
 \begin{equation}
 \chi^{(s)}(q,\omega)=\frac{\chi^{(0)}(q,\omega+{\text{i}}\gamma)}
 {1-\frac{{\text{i}}\gamma}{ \omega+{\text{i}}\gamma}
 \left[1-\frac{\chi^{(0)}(q,{\omega+{\text{i}}\gamma})}
 {\chi^{(0)}(q)}\right]}\ .
 \label{chi(s)}
 \end{equation}
$\chi^{(0)}(q,{\omega+{\text{i}}\gamma})$ is the susceptibility for
non-interacting carriers scattering off the disorder.  We use the
memory function formalism \cite {Mori} to calculate the scattering rate
$\gamma$.  This is related to the mobility $\mu$ by the Drude
expression $\mu=e/{m^\star\gamma}$.  At zero temperature in the
diffusive regime $\chi^{(s)}(q,\omega)$ is given by
$\lim_{\omega,q\rightarrow0}\chi^{(s)}(q,\omega)= (2m^\star)/(\pi
k_F\hbar^2)({\cal{D}}q^2)/({\cal{D}}q^2+\text{i}\omega)$, where
${\cal{D}}=v_F^2/\gamma$ is the diffusion constant.  In the limit when
$\gamma$ goes to infinity the system becomes non-diffusive. This
represents a localized phase \cite{TN1}.

In the memory function formalism $\gamma$  is expressed in terms of the 
non-linear equation \cite{Gotze}
 \begin{eqnarray}
 {\rm i}\gamma&=&-\frac{1}{2m^\star n_c}\sum_qq^2\left[
 \langle|U_{\text{imp}}(q)|^2\rangle+\langle|W_{\text{surf}}(q)|^2
 \rangle \right]\nonumber \\
 &&\times\left( \frac{\tilde\chi(q)}{\chi^{(0)}(q)}\right)^2
 \Phi(q,{\rm i}\gamma)\\
 \Phi(q,{\rm i}\gamma)&=&\frac{\phi_0(q,{\rm i}\gamma)}{1+{\rm i}\gamma
 \phi_0(q,{\rm i}\gamma)/\chi^{(0)}(q)}\ .
 \label{gamma}
 \end{eqnarray}
$\Phi(q,{\rm i}\gamma)$ is the relaxation spectrum for non-interacting
carriers that scatter off the disorder with scattering rate $\gamma$.
It is expressed in terms of $\phi_0(q,{\rm i}\gamma)=(1/{\rm
i}\gamma)\left[\chi^{(0)}(q,{\rm i}\gamma)-\chi^{(0)}(q)\right]$ which
is the relaxation function for non-interacting carriers with lifetime
$\gamma^{-1}$.  The static response function
$\tilde\chi(q)={\chi^{(0)}(q)}/\{1+V(q)[1-G(q)]\chi^{(0)}(q)\}$ 
includes the correlations between the carriers for the disordered
system.  $U_{\text{imp}}(q)=[(2\pi e^2)/(\epsilon q)]\exp(-qd)F_i(q)$
is the impurity potential for the monovalent Coulomb impurities which
are in a layer separated from the electron or hole plane by a distance
$d$.  We use for the impurity form factor $F_i(q)$ Eq.\ 4.28 in
Ref.\ \cite{AFS}.  For electrons in the Si MOSFETs we also include
interface surface roughness scattering.  This is the term
$W_{\text{surf}}(q)=\sqrt{\pi}\Delta\Lambda\Gamma(q)
\text{exp}(-(q\Lambda)^2/8)$.  Values for the parameters
$\Lambda=0.37$nm and $\Delta=2.0$nm are taken from Si MOSFET data
\cite{Khaikin}.  For $\Gamma(q)$ we use the expression in
Ref.\ \cite{Ando}.  For GaAs surface roughness scattering is much
smaller and we set $W_{\text{surf}}(q)=0$.

In the STLS formalism the density-density correlation function
$\langle\delta\hat{n}(r,t)\delta\hat{n}(r',t)\rangle$ is approximated
by the non-linear product, $\delta n(r,t)$$\times$$
g(r-r')$$\times$$\delta n(r',t)$.  The $\delta{n}(r,t)$ are expectation
values and $g(r)$ is the pair-correlation function giving the
probability of finding electrons a distance $r$ apart.  Using
$g(r)=1+n_c^{-1}\int{\text d}^2{\mathbf q}\ \exp({\text{i}}{\mathbf
q.\mathbf r})[S(q)-1]$, this gives us a relation between the static
structure factor $S(q)$ and the local field factor $G(q)$,
\begin{equation} 
G(q)= -{1\over n_c}\int{{\rm d}^2 {\bf k}\over
(2\pi)^2} {({\bf q}\cdot{\bf k})\over q^2}{V({\bf k})\over
V({\bf q})} [S(\mid{\mathbf q}-{\mathbf k}\mid) -1].
\label{G(q)} 
\end{equation}

For a given $G(q)$ we can determine $\gamma$ from Eq.\ \ref{gamma}.
Then $S(q)=(n_c\pi)^{-1}\int_0^\infty{\text{d}}\omega\;{\text
Im}\chi(q,\omega)$ is calculated from the fluctuation-dissipation
theorem and
 \begin{equation}
 \chi(q,\omega)=\frac{\chi^{(s)}(q, 
 \omega)}{1+V(q)\left[1-G(q)\right]\chi^{(s)}(q,\omega)}\ .
 \label{chiSTLSgamma} 
 \end{equation} 
Equation \ref{G(q)} gives a new local field factor which can be used in
Eq.\ \ref{gamma}.  The process is repeated iteratively until there is
overall self-consistency with both $\gamma$ and $G(q)$.

Our variables are the carrier density $n_c$, the impurity density
$n_i$, and the distance separating the impurities from the electron or
hole plane $d$.  The Metal Insulator Transition is observed at carrier
densities that are relatively low by conventional semiconductor
standards so the Fermi temperature $T_F=E_F/k_B$ can be of the order of
a few degrees K.  We have solved the equations at finite temperatures
and determined the dependence of $\gamma$ on temperature up to $T\alt
T_F$.

 \begin{figure}
 \epsfxsize=8.0cm
 \epsffile{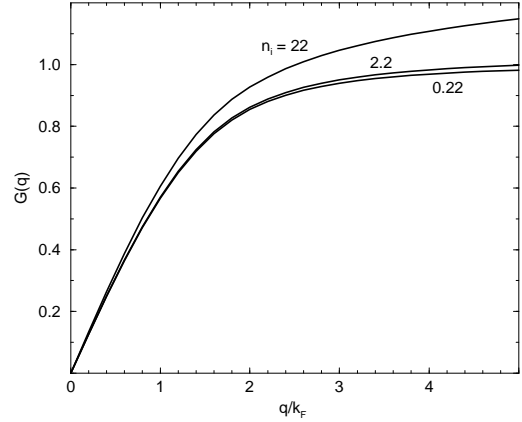}
  \caption[dummy3]{
Local-field-factor $G(q)$ for different impurity densities.
The impurities are separated from the carrier plane by
$d$$=$$5a_0^\star$.  Carrier density is
$n_c=35\times10^{10}$cm$^{-2}$.
 \label{Gq}}
\end{figure}

Figure \ref{Gq} shows the dependence of the local field factor $G(q)$
on the impurity concentration at zero temperature.  Increasing the
disorder enhances $G(q)$.  This is caused by the decrease in
$\chi^{(s)}$ as the scattering rate $\gamma$ gets bigger (Eq.\
\ref{chi(s)}).  Enhancing $G(q)$ weakens the effective interaction
between the carriers, and hence weakens the screening of the
carrier-impurity potential.  The net result of enhancing $G(q)$ is thus
to strengthen the effect of the disorder potential.  This in turn
further increases $\gamma$.  At a critical level of disorder this
non-linear feedback causes $\gamma$ to increase rapidly and leads to
localization of the carriers.

 \begin{figure}
 \epsfxsize=8.0cm
 \epsffile{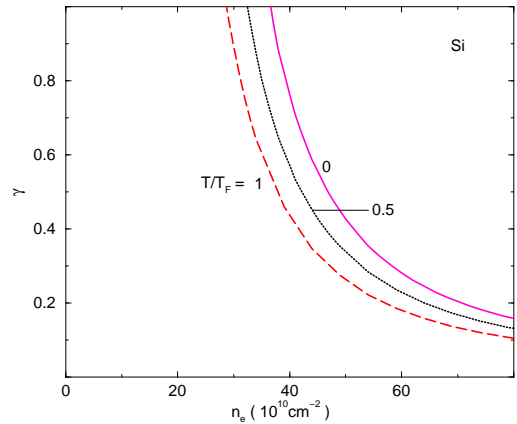}
  \caption[dummy3]{
Scattering rate $\gamma$ for Si as a function of electron
density for impurity density $0.5$$\times$$10^{11}$ cm$^{-2}$.
Separation of the impurity layer is $d$$=$$a_0^\star$.  Curves are for
temperatures $T$ (labels are in units of the Fermi temperature $T_F$
for density $n_c$).
 \label{gammavsneatT}}
\end{figure}

We see this non-linear behavior in Fig.\ \ref{gammavsneatT}.  At a
certain critical carrier density the scattering rate $\gamma$ starts to
increase rapidly.  The impurity density here is
$n_i=0.5$$\times$$10^{11}$cm$^{-2}$, with the impurities separated from
the carrier plane by distance $d=a_0^\star$ ($a_0^\star$ is the
effective Bohr radius).  The non-linear increase in $\gamma$ is due (i)
to the enhancement of the local field $G(q)$ which strengthens the
disorder potential, and (ii) to the rapid increase in the relaxation
spectrum $\Phi(q,{\rm i}\gamma)$ with $\gamma$.  Figure
\ref{gammavsneatT} also shows the dependence of $\gamma$ on
temperature.  The labels on the curves give the temperature in units of
$T_F$ for density $n_c$.  $\gamma$ diminishes with increasing
temperature.  This reflects the weakening of the correlations at finite
$T$.

 \begin{figure}
 \epsfxsize=8.0cm
 \epsffile{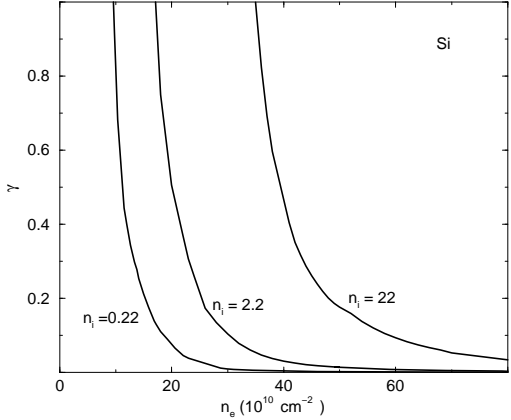}
  \caption[dummy3]{
Scattering rate $\gamma$ for Si at $T=0$ as a function of
electron density.  Curves are labelled for different impurity densities
$n_i$ (in units of $10^{11}$cm$^{-2}$). Impurity separation
$d$$=$$5a_0^\star$.
 \label{gammavsne}}
\end{figure}

In Figure \ref{gammavsne} we show the increase of $\gamma$ with
impurity density at $T=0$.  The curve labels give $n_i$ in units of
$10^{11}$cm$^{-2}$. The impurity separation is $d=5a_0^\star$.
As expected, increasing the impurity concentration has the effect of
increasing the scattering rate.  On the other hand, increasing the
carrier density has the opposite effect on $\gamma$ because the
correlations are reduced.  However the non-linear effects from $n_c$
are much stronger than the non-linear effects from $n_i$.

There exists no first order localization-delocalization transition in
two-dimensions.   In the metallic phase the mean-free-path is large
$\ell k_F\gg1$ while in the localized phase $\ell k_{F}$ should be
$\alt1$. We take the localization boundary to be the point where $\ell
k_F=1$.  Table I gives for different temperatures the critical electron
density at which $\ell k_F$ passes through unity.  When the system is
at finite temperature the correlations are reduced and one needs to go
to slightly lower carrier densities before localization can be
achieved.

\begin{figure}
 \epsfxsize=8.0cm
 \epsffile{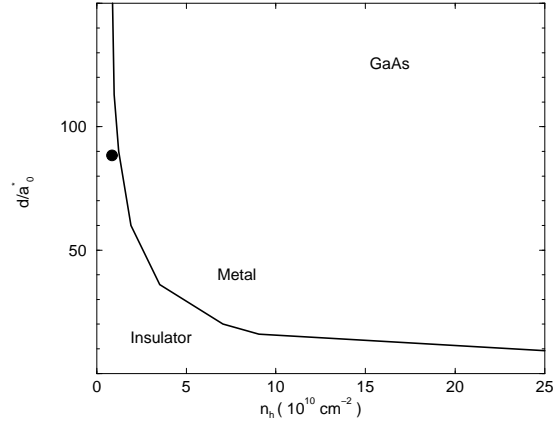}
  \caption[dummy3]{
Phase diagram for holes in GaAs. Critical hole density $n_h$ at
which $\gamma=1$ as a function of impurity layer separation $d$.
Impurity density is $n_i=22$$\times$$10^{11}$cm$^{-2}$.  The
experimental data point for GaAs is from Ref.\ \cite{simmons}.
 \label{dvsrs}}
\end{figure}

Table I shows the temperature dependence is quite small and so we give
Metal-Insulator phase diagrams for $T=0$.  The dependence of the hole
density at the transition on the impurity layer separation $d$ for GaAs
is shown in Fig.\ \ref{dvsrs}.  The impurity density is fixed at
$n_i=0.5$$\times$$10^{11}$cm$^{-2}$.  We compare our results with the
observation of the transition  in Ref.\ \cite{simmons} in a GaAs
sample.  Our curve is consistent with this measurement.  The phase
boundary is sensitive to $d$ because of the exponential factor in
$U_{\text{imp}}(q)$.  This suppresses carrier-impurity scattering for
short wavelengths $q$$\gg$$d$.  For the residual long wavelength
scattering the correlations are always weak.  Thus when $d$ is large
the correlations play a relatively less important role in the
localization.

 \begin{figure}
 \epsfxsize=8.0cm
 \epsffile{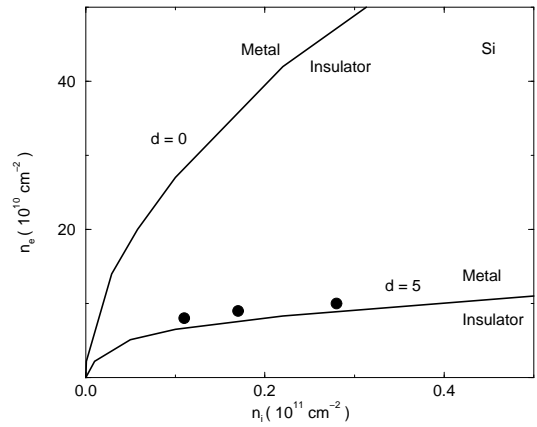}
  \caption[dummy3]{
Phase diagram for electrons in Si.  Critical carrier density at
which $\gamma=1$ as a function of impurity density for separations
$d$$=$$0$ and $5a_0^\star$.  The experimental data points are taken
from Ref.\ \cite{pudalov}.
 \label{nivsrs}}
\end{figure}

The phase diagram in Figure \ref{nivsrs} plots the critical electron
density in Si at the transition as a function of impurity density
$n_i$.  The impurity layer separations are $d=0$ and $5a_0^\star$.  We
also show experimental data points for the position of the
Metal-Insulator transition in Si \cite{pudalov}.  Again our predicted
phase boundary agrees with these observations.  On our curve the
electron-impurity scattering becomes weaker as the critical electron
density decreases.  Thus at the transition point the electron mobility
will increase as the electron density decreases.  This is consistent
with recent observations \cite{pudalov}.

We conclude that correlations and impurity scattering do mutually
affect the localization transition.  Finite temperatures tend to
suppress correlations and this slightly reduces the critical carrier
density for localization to occur by an amount which becomes
significant for temperatures of the order $T_F$.

\acknowledgements 

We acknowledge financial support from an Australian Research
Council Grant (JST) and an INFM/Forum Italian fellowship (LL).

\ \\
\begin{center}
\ 
\begin{tabular}{||c||c|c|c|c||}
\hline
&&&&\\
$T$ (K) &$0$\ &$1$\ &$10$\ &$20$\\
\hline
&&&&\\
\ $n_e$ (cm$^{-2}$)\ \ &\ \ $37\times10^{10}$\ \ \ &
\ \ $37\times10^{10}$\ \ \ &\ \ $32\times10^{10}$\ \ \ &
\ \ $29\times10^{10}$\ \ \\
&&&&\\
\hline
\end{tabular}

\ \\ \ \\
Table I

\noindent
Dependence on temperature of the critical carrier density for
localization in Si.\\  Impurity density
$n_i$$=$$0.5$$\times$$10^{11}$cm$^{-2}$.  Impurities are embedded in
the carrier plane.  \end{center}
\narrowtext

\end{document}